\documentclass[pra,10pt,tightenlines,twocolumn,superscriptaddress,dvipsnames]{revtex4-1}
\usepackage{amsmath,amssymb,amsfonts,xcolor,graphicx,float}
\usepackage{multirow}
\usepackage{bm}
\usepackage{blindtext}
\usepackage{comment}
\usepackage{color, colortbl}
\newcommand\change[1]{\textcolor{black}{{#1}}}

\begin{document}
\title{Quantum topology identification  with deep neural networks and quantum walks}
\author{Yurui Ming}%
\affiliation{Centre for Artificial Intelligence, School of Computer Science, University of Technology Sydney, Sydney, Australia}%
\author{Chin-Teng Lin}%
\affiliation{Centre for Artificial Intelligence, School of Computer Science, University of Technology Sydney, Sydney, Australia}%
\author{Stephen D. Bartlett}%
\affiliation{Centre for Engineered Quantum Systems, School of Physics, The University of Sydney, Sydney, Australia}%
\author{Wei-Wei Zhang}%
\email{ww.zhang@sydney.edu.au}
\affiliation{Centre for Engineered Quantum Systems, School of Physics, The University of Sydney, Sydney, Australia}%

\date{\today}
\begin{abstract}
Topologically ordered materials may serve as a platform for new quantum technologies such as fault-tolerant quantum computers.  To fulfil this promise, efficient and general methods are needed to discover and classify new topological phases of matter.  We demonstrate that deep neural networks augmented with external memory can use the density profiles formed in quantum walks to efficiently identify properties of a topological phase as well as phase transitions.  On a trial topological ordered model, our method's accuracy of topological phase identification reaches $97.4\%$, and is shown to be robust to noise on the data. Furthermore, we demonstrate that our trained DNN is able to identify topological phases of a perturbed model and predict the corresponding shift of topological phase transitions without learning any information about the perturbations in advance. 
These results demonstrate that our approach is generally applicable and may be used to identify a variety of quantum topological materials.
\end{abstract}
\maketitle

\textbf{Introduction}  

The properties of topological quantum materials have been the subject of intense interest in recent years, due to their paradigm-changing implications for condensed matter physics~\cite{Moore_2010,Hasan_2010,Ryu_2010,Qi_2011} and  
potential applications to new technologies. 
The electric conductivity of topological materials such as topological insulators has potential applications for magnetoelectric devices with higher efficiency and lower energy consumption~\cite{Li_2014,Ando_2014,DC_2018}.  In addition, topological materials can support anyonic quasiparticle excitations, with exotic statistics under braiding transformations that may enable fault-tolerant quantum computing~\cite{Nayak_2008, Field_2018}.
The topological ordering of quantum materials can be characterised with quantised, nonlocal topological invariants, such as the Chern number of the quantum Hall effect. These invariants determine all of the key topological properties of quantum systems, such as the number of topological edge states and the types of anyonic excitations in topological materials. The discovery and characterisation of novel topological quantum materials requires a general and efficient method to identify these topological invariants using experimentally accessible properties. For bulk systems of topological insulators, these can often be inferred from the existence of edge states~\cite{Hasan_2010,Wu_2016}, or particle dynamics, such as the anomalous velocities obtained by wave packets under applied forces~\cite{Price_2012,Duca_2015}, and quantum walks~\cite{Kitagawa_2010,Kitagawa_2012,Cardano_2016,Zhang_2017deco,Zhang_2017,zhang2017detecting,Sun_2018}. However, despite the considerable theoretical progress in developing classification methods for topological phases, we still lack a universal automatic method for the discovery and characterisation of new materials.

Here we propose and test a universal automated method for identifying topological phases of quantum materials, combining quantum walks to probe the phase and a deep neural network (DNN) to analyse the evolution.   Using the particle density profiles formed during a particle's evolution driven by the system's Hamiltonian, we demonstrate that a novel DNN with external memory is able to identify the topological phases and phase transitions for a two-dimensional lattice model with spin-orbit coupling.  Our method demonstrates high identification accuracy of $97.4\%$, and is shown to be robust to noise on the input data.  Finally, although we train our model using data from a specific two-dimensional spin-orbit lattice Hamiltonian, we demonstrate that our method is able to classify the phases of a perturbed model with high accuracy, without any details about the perturbation. As such, our results demonstrate that quantum walks and DNN are a powerful and generic tool for the efficient discovery and analysis of novel topological quantum systems, and therefore the design of robust quantum technologies.

\textbf{Results} 
 
 \textit{\textbf{Continuous-time quantum walks in topological quantum systems.}}  
 The coherent dynamics of particles, with motion dependent on an internal degree of freedom such as spin, are described as quantum walks.  Along with providing a tool for building quantum algorithms, quantum walks also provide a platform to simulate and analyse complex physical systems~\cite{Venegas_2012,portugal2013quantum}. There are two types of quantum walks: discrete-- and continuous--time quantum walks, where the main difference is the timing used to apply corresponding evolution operators. In the case of discrete--time quantum walks, the corresponding evolution operator of the system is applied only in discrete time steps, while in the continuous--time quantum walk case, the evolution operator is  applied continuously. 
   Discrete-time quantum walks have been successfully used to study topological properties of a quantum system.  Specifically,  the experimental observation of particle localisation at the boundary between materials possessing different topological ordering and its robustness to the defects have been used to prove the existence of topologically protected edge modes~\cite{Kitagawa_2010,Kitagawa_2012,Cardano_2016,Zhang_2017deco,Flurin_PRX_2017,xiao2017observation}.   Furthermore,  the moments of the probability distribution for the walker's position after many steps is an experimental signature of a topological quantum phase transition in one-dimensional quantum walks~\cite{Cardano_2016}.
 In contrast to discrete-time quantum walks, which require pulsed control over the system,  continuous-time quantum walks (CTQW) can arise directly in free Hamiltonian systems such as two-dimensional spin-orbit lattice models.  These CTQW have been shown to reveal topological phase transitions~\cite{Zhang_2017}, a fact supported by recent experiments~\cite{Sun_2018}.  In such CTQW, the resulting density profile of an initially localized particle is expected to contain a wealth of information to identify the topological order of the underlying quantum system, provided one can extract this information efficiently.

\begin{figure}[h]
\centering
\includegraphics[width=0.95\linewidth]{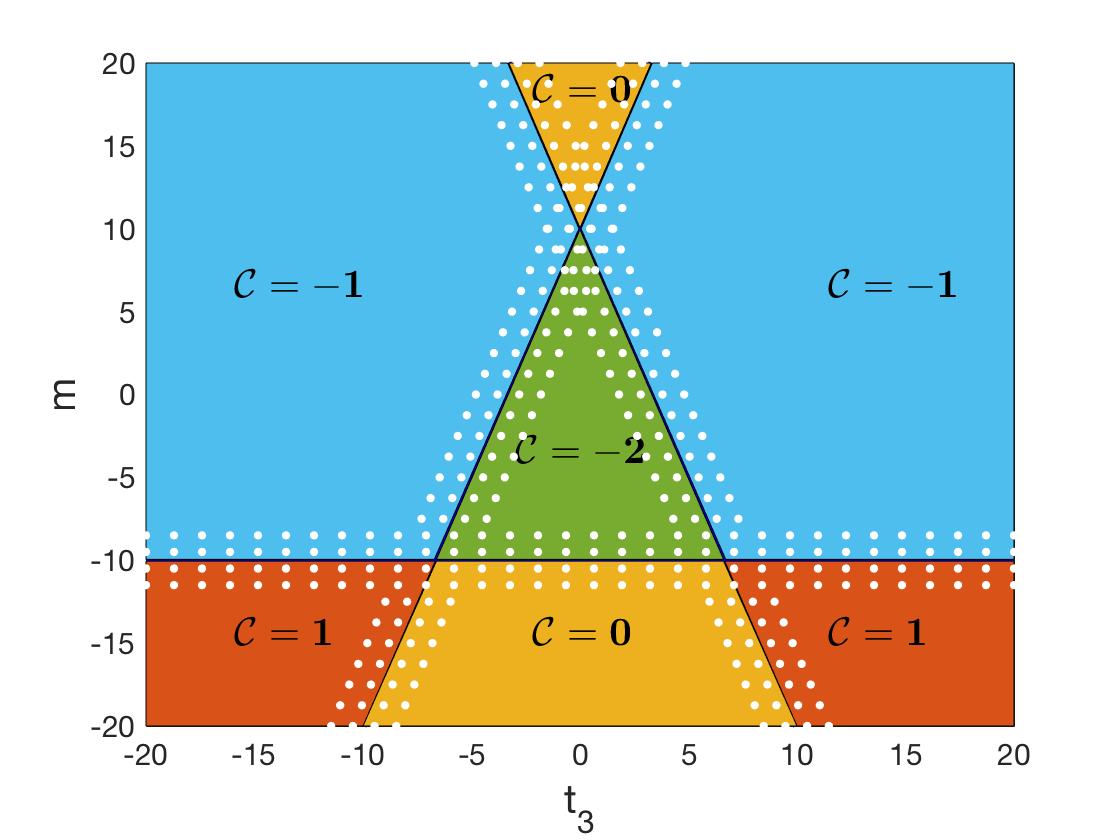}
\caption{The parameter space of the Hamiltonian of Eq.~(\ref{Eq:hamiltonian}) with fixed $t_{1x}=t_{1y}=1$, $t_2=5$, used for the generation of our dataset.  The coloured regions represent the ``whole'' areas, and the dotted regions are the ``transition'' areas, as detailed in Table~\ref{acc_table}. 
A parameter space containing the phase $\mathcal{C}=+2$ is similar, with this phase located at the same region of $\mathcal{C}=-2$, obtained by flipping the sign of $t_{1y}$ and keeping the other parameters the same. 
\label{phase_diagram}}
\end{figure}

 \begin{figure*}[h]
\includegraphics[width=0.88\textwidth]{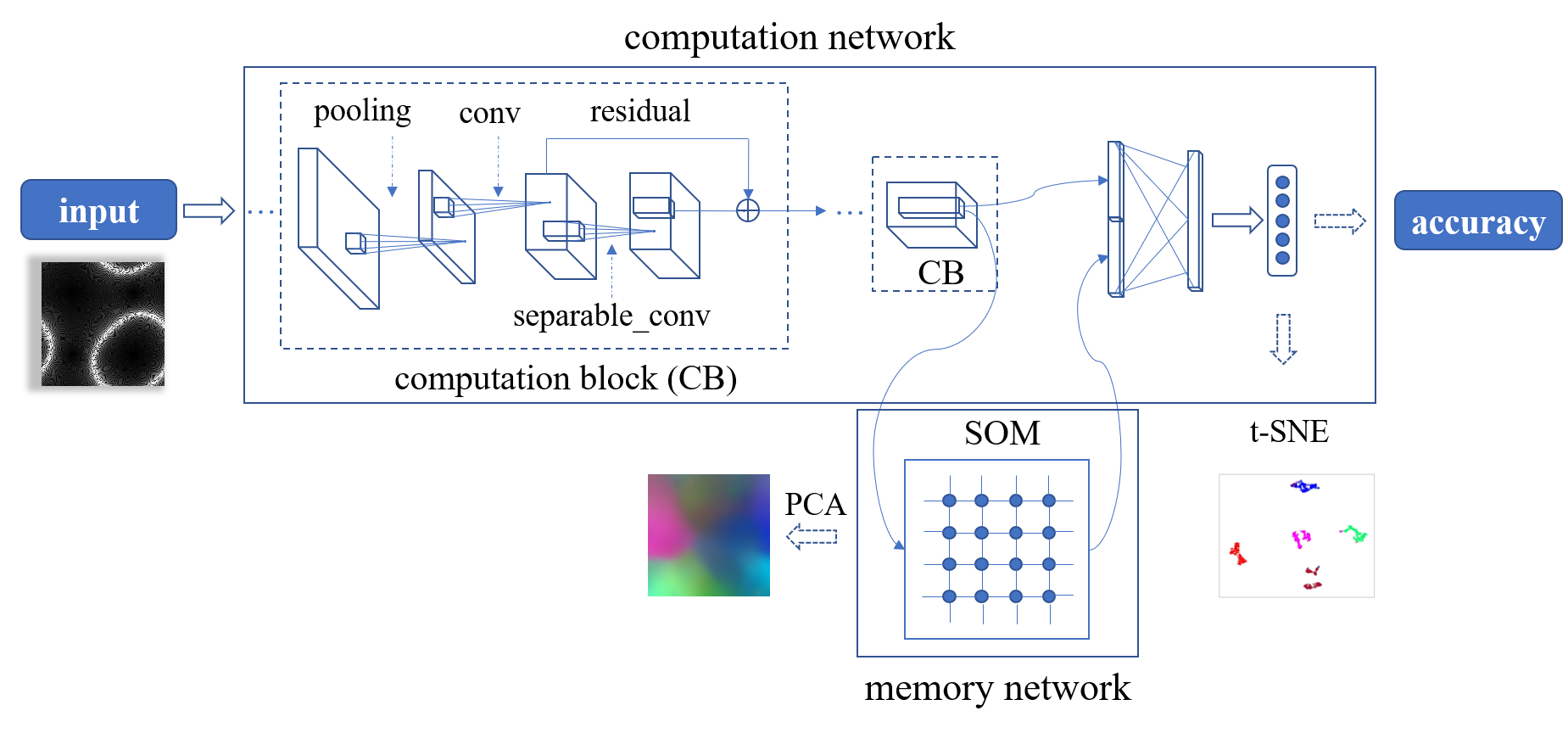}\,\,
\caption{Network architecture: the computation network is constructed from six computation blocks of a supervised learning paradigm, the memory network is of an unsupervised learning paradigm. ``SOM'' represents for self-organising map,   ``conv'' represents for convolution. The input of our DNN is the density profiles and the outputs are PCA, t-SNE and statistical accuracy. }
\label{DNN}
\end{figure*}

\begin{figure*}[h]
\includegraphics[width=0.76\textwidth]{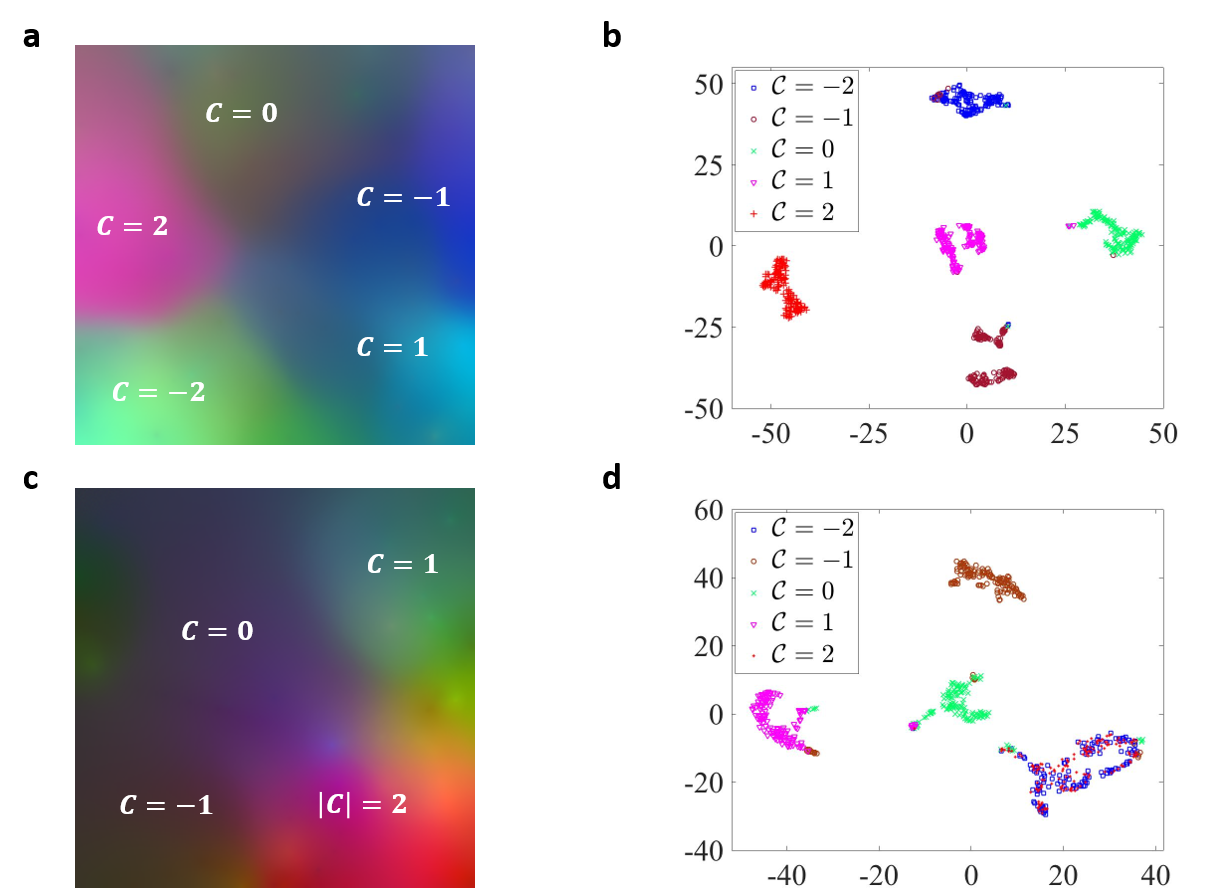}
\caption{ PCA of memory based on momentum (a) and position (c) density profiles, which illustrate the self-organised clusters formed in memory  during our training process and shows the clustering of the input data. The size of PCA is same as the size of the memory in our DNN ($256\times 256$). 
 The  RGB colour  is obtained by projecting the 32 dimensional vector of each memory pixel into a 3 channel colour representation.  In our experiments, the scree plots indicate that the first three components explain around $80\%$ of variance, and an ``elbow", the cutting-off point, appears at the third principal component. This justifies our choice of first three principal components in our experiments. (b) and (d) are classification visualisations of momentum and position samples via t-SNE, which is a projection from the 5 dimensional DNN output vector into the  location indices of a two-dimensional space.}
\label{momentum-position}
\end{figure*}

\begin{figure*}[h]
\includegraphics[width=0.85\textwidth]{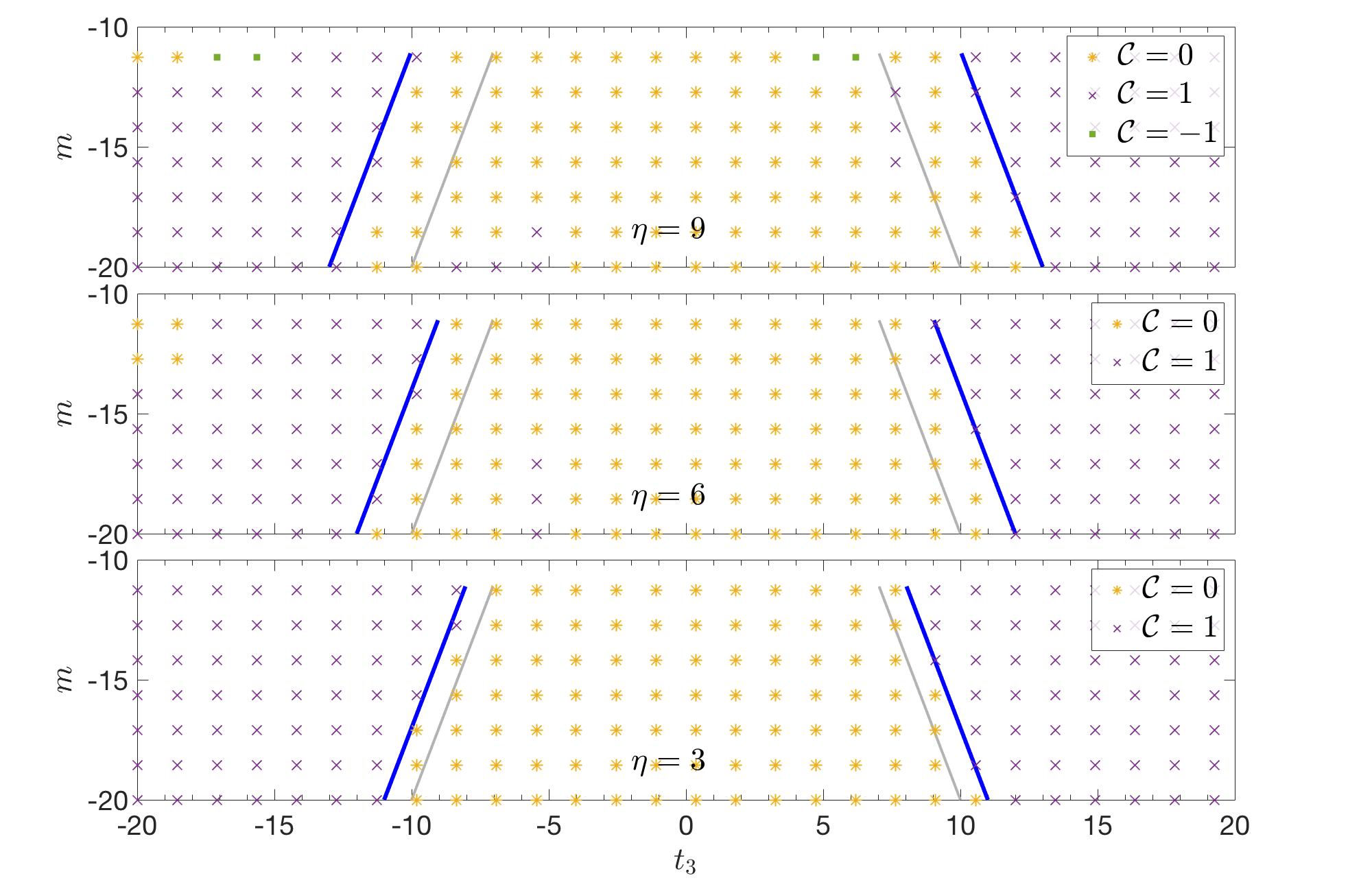}
\caption{\change{Topological phase boundary shift of perturbed systems with perturbation strengths $\eta=\{3,6,9\}$ (from bottom to top) identified with the trained DNN. The Chern number of the perturbed systems identified with our trained DNN  are represented with  the symbols  ``{\large\color{BurntOrange}$*$}, {\small\color{Plum}$\times$},~{\color{YellowGreen}\tiny$\blacksquare$}" for $\mathcal{C}=0, 1, -1$ respectively, and the coordinates of the symbols represent the system's  parameters $\{t_3,~m\}$. The thin grey line is the phase boundary of unperturbed system ($\eta=0$), the blue line is the theoretical prediction of phase boundary for the perturbed systems. The horizontal displacement of the blue lines from the grey line indicates the topological phase boundary shift of the corresponding perturbed system.}}
\label{boundary_shift}
\end{figure*}

 In this work, we consider the topological phases of  a parameterised Hamiltonian on a two-dimensional lattice ($599\times 599$ in our simulation).  Following Ref.~\cite{Zhang_2017}, we use a continuous-time quantum walk (CTQW) for a initially localised  spin up particle under this Hamiltonian, where the behaviour of the distribution of the quantum state after long time evolution provides a signature of the topological phase.  We will investigate the use of both the particle's spatial as well as its momentum density profiles, marginalizing over the particles internal state. 
Specifically, we consider the two-dimensional spin-orbit lattice Hamiltonian~\cite{Sticlet_2012,Asboth_2016,Zhang_2017,Sun_2018}, as described in the \textit{Methods}.   We use this model to test our method for topological phase identification because the topological invariant (Chern number) of this system is easily calculated,  allowing us to check the accuracy of our method.    This Hamiltonian supports five distinct topological phases, labelled by the Chern number $\mathcal{C}\in \left\{0,\pm1,\pm2\right\}$, determined by the coupling parameters in this Hamiltonian, as shown in Fig.~(\ref{phase_diagram}).

The density profile is strongly dependent on the system's topology, and can be used as a diagnostic of topological phases, and the phase transitions between them, as discussed in Refs.~\cite{Zhang_2017,Sun_2018}.  From these previous studies, good signatures for topological phase identification are the central features of the position distribution and the ring pattern of the momentum distribution, which reveal that the Hamiltonian localizes in a nontrivial topological phase with Chern number $\mathcal{C}=\pm1$.  However, these previous analyses are based on approximations, and we do not have a general method to analyse the density profiles for topological phases associated with other Chern numbers. 

 \textit{\textbf{Learning topological phases using a deep neural network.}} 
Machine learning can determine the underlying characteristics of a physical system even without prior human knowledge~\cite{Schmidt_2009}.  Deep learning, a subset of machine learning  which represents the data as a nested hierarchy of concepts, provide great capability and adaptability in this regard~\cite{goodfellow2016deep}. Each concept is defined in relation to simpler concepts, and more abstract representations are computed in terms of less abstract ones.  Deep learning has achieved breakthroughs across many  applications~\cite{goodfellow2016deep,krizhevsky2012imagenet,esteva2017dermatologist,shallue2018identifying},  indicating its potential benefit in the analysis of many different quantum problems~\cite{Cai_2015,Schuld_2015,Dunjko_2016,Biamonte_2017,Mott_2017,Broecker_2017,Carrasquilla_2017,Zhang_2017qml,Zhang_2018,Choo_2018, Lu_2018, arrazola2018machine,Caio_2019,Rem_2018,Mehta2019, Sarma_2019,Schuld_2019,Rodriguez2019}.  Inspired by the hierarchical bio-structures in visual systems~\cite{hubel2012david}, deep neural networks (DNN) can automatically extract the most suitable representations from input data and make accurate predictions. 
Generally speaking, during the end-to-end learning process, the representations of data will automatically emerge rather than being discovered or manually crafted~\cite{lecun2015deep}.

We will apply DNN to the problem of topological identification by providing the network with the density profiles from a CTQW as input.  As described above, the density profiles contain a wealth of information about the topological phase of the system, but identifying which features are important is challenging, especially for higher order phases.  A DNN  with external memory has the capacity to solve complex structural tasks that are challenging to stand-alone neural networks, and has shown the ability to answer synthetic questions designed to emulate reasoning and inference problems~\cite{Graves_2016}.  The architecture of our deep neural network is shown in Fig.~(\ref{DNN}), which consists of multiple computation blocks (CB) and fully connected layers (computation network), as well as an external memory coupled to the last convolutional layer (memory network). The computation network is of a supervised-learning paradigm and the memory network is of an unsupervised-learning paradigm.  Supervised and unsupervised paradigms each have their own advantages in classification problems, as introduced in Ref.~\cite{goodfellow2016deep}, and they are jointly trained during the process in our experiments.

Our experiment consists of three steps: data preparation, neural network training and validation, and testing.  
The data preparation stage is based on numerically simulations of CTQW with different Hamiltonian parameters, and is described in the \textit{Methods}.  The data corresponding to different topological phases is randomised and split into three sets with the ratio $0.8:0.1:0.1$  for training, validation and testing respectively.  \change{Validation is integrated to the iterative training process to prevent overfitting.} Details of Neural architecture evaluation and naive baseline are given in the Appendix.
The prepared data is reused three times to evaluate the network.  As the performance indicator for the corresponding prepared data, the accuracy in our results is the average over  the three independent randomisation sets.

We  analyse the outcome of our experiments using the principal component analysis (PCA) of memory, a t-distributed stochastic neighbour embedding (t-SNE) of the computation network output, and the statistical accuracy of the test. Both the PCA and the t-SNE are visualisation results, and the accuracy is a statistical evaluation. 
The t-SNE shows the topological classification of input data corresponding to different Chern numbers.   The PCA demonstrates how the input data is clustered according to its correlation by self-organisation, which distinguishes the different topological phases of the input data.  The accuracy  represents the fraction of test data that is correctly identified (by comparing with the analytical solution).

\begin{table*}[h]
\caption{The statistical accuracy for the topological identification using our DNN. The accuracy  is obtained by averaging over three randomised data sets, where every data set is trained three times. (Note: if the data denoted with an asterisk (*) for the position data sets $\mathcal{C}=\pm2$ are excluded, we obtain a higher overall accuracy as denoted by a dagger ($^\dagger$) compared with the overall accuracy including this data.)}
\centering
\begin{tabular}{c|c|c|c|c|c|c|c}
\hline\hline
 \multicolumn{7}{c}{The statistical accuracy with ideal input data}\\
\hline\hline
 \multicolumn{2}{c}{Density Profile Data} &  \multicolumn{5}{|c|}{$\mathcal{C}$}  & \multirow{2}{*}{Overall}\\
\cline{1-7}
Phase Diagram Area  & Measurement  Domain & $-2$ & $-1$ & $0$ & $1$ & $2$  &  \\
\hline
  \multirow{2}{*}{Whole} & Momentum &0.974&0.961 & 0.969 & 0.969 &1.000 & 0.974 \\
                                    & Position & 0.703*&0.935 &0.937 & 0.975 & 0.241* & 0.761 ($0.949^\dagger$)\\
\hline
\multirow{2}{*}{Transition}& Momentum&0.946&0.914 & 0.965 & 0.968 &0.999 & 0.958 \\
                                        & Position & 0.913&0.908 &0.915 & 0.967 & 0.985 & 0.938\\
\hline\hline
 \multicolumn{7}{c}{The statistical accuracy with noisy input data}\\
\hline\hline
 \multicolumn{2}{c}{Density Profile Data} &  \multicolumn{5}{|c|}{$\mathcal{C}$}  & \multirow{2}{*}{Overall}\\
\cline{1-7}
Phase Diagram Area  & Measurement  Domain & $-2$ & $-1$ & $0$ & $1$ & $2$  &  \\
\hline
  \multirow{2}{*}{Whole} & Momentum &0.952 &0.953 & 0.969 & 0.987 &1.000 & 0.972 \\
                                    & Position & 0.536* &0.934 & 0.913 & 0.963 & 0.420* & 0.756 ($0.937^\dagger$)\\
\hline
\multirow{2}{*}{Transition}& Momentum&0.924&0.949 & 0.968 & 0.927 &1.000 & 0.954 \\
                                        & Position & 0.917&0.898 &0.918 & 0.936 & 0.973 & 0.929\\
\hline
\end{tabular}

\label{acc_table}
\end{table*}

The PCA and t-SNE based on the data---the density profiles in momentum and position space---are shown in Fig.~(\ref{momentum-position}), 
where the DNN identification forms separated clusters associated with the  topological phases of our model Hamiltonian system.
For the momentum space data, the identification clearly reveals five clusters corresponding to each of the distinct topological phases of the Hamiltonian.  For the position space data  covering the whole phase diagram, only four clusters are identified and the topological phases corresponding to Chern numbers $\mathcal{C}=\pm2$ are not distinguished based on this data.

The statistical accuracy of our test, i.e., the ratio between the number of testing samples classified into correct topological phases and the total number of testing samples, is shown in Table~\ref{acc_table}.  When based on momentum space density profiles, we obtain a very high accuracy for data covering both the whole phase diagram region,  as well as for a restriction to the region around the phase transition ($97.4\%$ and $95.8\%$ respectively). 
 Position space density profiles lead to identification with relatively lower accuracy for  the whole phase diagram, $76.1\%$, remaining high for the phase transitions regions  $93.8\%$.  The reduction in accuracy for the whole phase diagram is primarily because our DNN is unable to distinguish the phases $\mathcal{C}=\pm2$.  By excluding the data for $|\mathcal{C}|=2$, the accuracy obtained with data from the whole phase diagram reaches $94.9\%$. The low accuracies for distinguishing $\mathcal{C}=\pm 2$ in this case may potentially be an affect of our parameterisation of phase space: the variation of our chosen hyper-parameter manifold in FBZ for Hamiltonians with $\mathcal{C}=0,\pm1,-2$ are a continuous process, while the $\mathcal{C}=+2$ region is accessed through a discrete change; see the \textit{Methods} details of the data generation. The relatively small region for $|\mathcal{C}|=2$ in the phase diagram of this model also potentially restricts the learning ability of DNN.

\textit{Noisy data as input.} 
Quantum walks on engineered topological quantum materials have been realised in different physical platforms including photonics systems~\cite{zhang2017detecting,xiao2017observation, chen2018observation,Kitagawa_2012} and cold atoms~\cite{Sun_2018}, amongst others.  For our method to be useful on experimental data, it must be robust to noise.  Here, we test the performance of our method with noisy input data for our trained DNN.  We add Gaussian noise to our simulated data, at a level comparable with current experimental techniques in optical systems~\cite{xiao2017observation, chen2018observation} and cold atoms systems~\cite{Sun_2018, Wu_2016,robens2017high}; details are discussed in the \textit{Methods}.  In these tests, the accuracy statistics for topological phase identification shows limited degradation as indicated in Table~\ref{acc_table}.  Using momentum density profiles, the accuracy  decreases by only $0.3\%$ on average,  and this decrease could potentially be offset by increasing the size of the network.

\textit{\textbf{General applicability of the method.}} 
 As we now show, our DNN trained with the data from CTQWs governed by a known model is also able to identify the topology of a perturbed model without additional information or further training to learn the perturbation.  Hereafter, we refer the DNN after training on the unperturbed model as our ``trained DNN''.  In our test, the perturbed model is obtained by adding an additional term to our training Hamiltonian; see the \textit{Methods} for details.  As the Chern number for the perturbed model can still be calculated analytically, we are able to test the accuracy of our trained DNN to identify the topology of the perturbed model.

We generate three sets of momentum density profiles using the perturbed Hamiltonian with three different perturbation strengths $\eta=\{3, 6, 9\}$.  Our trained DNN is able to identify the topology of the perturbed Hamiltonian with an averaged accuracy $93.88\%$, with $\{97.96\%, 93.88\%, 89.80\%\}$ for $\eta=\{3, 6, 9\}$ respectively, where the accuracy decreases while increasing the perturbation strength.

 Furthermore, to demonstrate that our trained DNN is able to detect changes to the topological system caused by the perturbation, we show that it can identify the location of the topological phase transitions and how these locations shift depending on the perturbation.  Our trained DNN reveals that, while increasing the perturbation strength $\eta$, the phase boundary between $\mathcal{C}=1$ and $\mathcal{C}=0$ shifts in the direction of increasing magnitude of $t_3$, that is, the area of the $\mathcal{C}=0$ phase region is increasing as a function of the perturbation strength. Specifically, our trained DNN predictions for the phase transition shifts are $\Delta_\text{DNN}=\left\{1.005, 1.746, 2.686\right\}$ for $\eta=\{3, 6, 9\}$, which are close to  the theoretical analysis for the corresponding shifts $\Delta=\left\{1, 2, 3\right\}$.  We note that we classify phases using a grid of discrete points in parameter space, and that this discretisation accounts for a considerable uncertainty in our identified phase boundaries, comparable with the error in the estimates.  Further details are give in the \textit{Methods}.

\textbf{Conclusion and outlook.}

We have demonstrated a universal automatic method for the identification of distinct topological phases of quantum materials,  and the related perturbed models.  Our simulated experimental results show that the combination of the particle's density profile from a CTQW and DNN augmented with external memory is a reliable and  efficient method to identify topological phases and phase transitions in our trial system, even for the high order $\mathcal{C}=\pm2$ and noisy data.  We have also demonstrated the generality of this method, by using our trained DNN to classify the topological properties of a perturbed system without any knowledge of the perturbation. 

For the purpose of engineering novel topological systems using our method, we could use zero-shot learning methods, which aim to recognise objects whose instances may not have been seen during training~\cite{xian2018zero}. By integrating the zero-shot learning into our DNN, the design and identification of novel topological phases will be possible.

\textbf{Methods}

Here we present the trial topological Hamiltonian system, and describe the generation of a particle's density profile as used as the input data for our DNN.  The perturbed model, which we use to assess the generality of our method, is detailed as well.  We also provide the details of the architecture of our DNN. 

\textit{The topological system in our simulated experiments.}
The two-dimensional spin-orbit lattice Hamiltonian we consider here is~\cite{Sticlet_2012,Asboth_2016,Zhang_2017,Sun_2018}
\begin{align}
\hat{{H}}&=\sum_{x,y}\Big[c_{x,y}^{\dagger}\frac{m}{2}\hat{\sigma}_3c_{x,y}+c_{x+1,y}^{\dagger}(t_{1x}\hat{\sigma}_1-\text{i}\frac{3}{4}t_3\hat{\sigma}_3)c_{x,y}
\nonumber\\
&+c_{x,y+1}^{\dagger}(t_{1y}\hat{\sigma}_2-\text{i}\frac{3}{4}t_3\hat{\sigma}_3)c_{x,y}+c_{x+1,y+1}^{\dagger}t_2\hat{\sigma}_3c_{x,y}
\nonumber\\
&+h.c. \Big]\nonumber\\
&=\sum_{k_x,k_y}\vec{h}\cdot\vec{\sigma}\left | k_x,k_y\right\rangle\left\langle k_x,k_y\right |\,,
\label{Eq:hamiltonian_mom}
\end{align}
using $\{m,t_{1x},t_{1y},t_2,t_3\}$ as the coupling parameters, $i\in\{1,2,3\}$,  $\vec{\bf{\sigma}}=\{\hat{\sigma}_1,\hat{\sigma}_2,\hat{\sigma}_3\}$ as the Pauli operators and $\vec{h}=\left(h_1,h_2,h_3\right)$.  The last line of
 Eq.~(\ref{Eq:hamiltonian_mom}) is obtained by using translation invariance
and the Fourier Transformation $\{\left | k_x\right\rangle=\frac{1}{\sqrt{2\pi }}\sum_{x}e^{-ixk_x}\left | x\right\rangle, \left | k_y\right\rangle=\frac{1}{\sqrt{2\pi }}\sum_{y}e^{-iyk_y}\left | y\right\rangle\}$, the $2 \times 2$ block-diagonalized   Hamiltonian in momentum space is
\begin{align}
\vec{h}\cdot\vec{\sigma}&=2t_{1x}\cos k_x\hat{\sigma}_1+2t_{1y}\cos k_y\hat{\sigma}_2
\nonumber\\
&+\left\{m+2t_2\cos\left(k_x+k_y\right)+\frac{3}{2}t_3\left(\sin k_x+\sin k_y\right)
\right\}\hat{\sigma}_3\,.
\label{Eq:hamiltonian}
\end{align}
This Hamiltonian supports the topological phases with  Chern numbers $\mathcal{C}\in \left\{0,\pm1, \pm2\right\}$.  We consider a parameter space given by varying the coupling parameters $m$ and $t_3$ while fixing all other parameters.  For example,  while fixing {$t_{1x}=t_{1y}=1$, $t_2=5$} the Hamiltonian supports $\mathcal{C}\in \left\{0,\pm1, -2\right\}$  and  while fixing {$t_{1x}=1$, $t_{1y}=-1$, $t_2=5$} the Hamiltonian supports $\mathcal{C}\in \left\{0,\pm1, 2\right\}$. 
 The definition of 
Chern number is  
\begin{equation}
\mathcal{C}=\frac{1}{4\pi}\int_\text{BZ}d^2{\bf{k}}\ \hat{h}\cdot\left(\partial_{k_x}\hat{h}\times\partial_{ky}\hat{h}\right)\,,
\label{Chern_number}
\end{equation}
with $\hat{h}=\vec{h}/|\vec{h}|$~\cite{Sticlet_2012}. 
The different topological phases labelled by Chern number $\mathcal{C}$, as a function of Hamiltonian parameters,  are shown in Fig.~(\ref{phase_diagram}).

 \textit{The formation of particle's density profile in both momentum and position spaces.} 
 In CTQW evolutions, a particle with spin up, initially localised in the centre of a two-dimensional lattice in position space,  spreads out and gradually occupies  a larger area of the lattice. Equivalently, the particle is initially uniformly distributed in momentum space and during the evolution the particle's components at every momenta oscillates between spin up and spin down components.  The particle's wave functions and probability distributions in both position and momentum spaces form a certain pattern which is closely related with the Hamiltonian.

 At evolution time $t$, the state of the particle initially  spin up  and localised at the centre of two-dimensional lattice is (setting $\hbar=1$)
\begin{align}
\left|\psi(t)\right\rangle
&=\sum_{\bm k} \left(\alpha_{\bm k\uparrow}\left|\uparrow\right\rangle+\alpha_{\bm k\downarrow}\left|\downarrow\right\rangle\right)\left|\bm k\right\rangle \nonumber\\
&=\sum_{\bm k}
\begin{pmatrix} \frac{h_3\left(-\text{i}\text{sin}\left(E_{\bm k}t\right)\right)}{E_{\bm k}}-\text{cos}\left(E_{\bm k}t\right)\cr 
\frac{\left(h_1+\text{i}h_2\right)\left(-\text{i}\text{sin}\left(E_{\bm k}t\right)\right)}{E_{\bm k}}\end{pmatrix}
|{\bm k}\rangle\, 
\label{Eq:finalstate}
\end{align}
where $E_{\bm k}=\sqrt{h_x^2+h_y^2+h_z^2}\neq0$ is the eigenenergy of system's Hamiltonian. When $E_{\bm k}=0$ we have $\alpha_{\bm k\uparrow}=1$ and $\alpha_{\bm k\downarrow}=0$, which is the case at Dirac point while the system is under topological phase transition. 
The particle's state represented in position space is the Fourier transform of the corresponding spin components.
 
From the expression of Eq.~(\ref{Eq:finalstate}) for particle's state at time $t$, the amplitude and the relative phase of both spin up and spin down components are closely related with the energy $E_{\bm k}$ and sensitive to the band gap of the system which is $\text{min}\{2E_{\bm k}\}$ as discussed in Ref.~\cite{Zhang_2017,Sun_2018}. The topological phase of the system characterised with Chern number is revealed by the band structure of the system. Therefore, the particle's density profile is a competitive candidate for the  topological detection, even for higher order phases.

Here,  we generate two sets of density profiles. One is the wave functions in momentum space and the other  is the probability distributions in position space. For the training of the neural network, we decompose the complex values of both spin up and spin down components into two real values and map the  amplitude and relative phase matrices into  image representation.  With this process, the input data set consists of the set of spatial or momentum distributions for the particle's final states.

\textit{Dataset generation for our deep neural network identifying the topology of quantum matters.}
Our system supports topological phases with $\mathcal{C}=\{0,\pm1,\pm2\}$ as described above. The diagram showing the distribution of Chern number $\mathcal{C}$ with respect to ${m,t_3}$ and fixed $t_{1x}=t_{1y}=1, t_2=5$ is shown in Fig.~(\ref{phase_diagram}), where the shaded area represents the parameter area for the dataset  labeled as ``whole''  and the dotted area represents the parameter area for the dataset labeled as  ``transition'' in our tables. The dataset for $\mathcal{C}=2$ is generated with the same $m,t_3,t_{1x},t_2$ as $\mathcal{C}=-2$, but with ${t_{1y}=-1}$. The sizes of our dataset generated for the whole phase diagram are $\{1449, 1478, 1486, 1488, 1449\}$ and for the phase transition area of the diagram are  $\{1575, 1506, 1474, 1408, 1575\}$  corresponding to $\mathcal{C}=\{-2, -1, 0, 1, 2\}$ respectively.  The conventional practice using DNN~\cite{krizhevsky2012imagenet} indicates this data size is sufficient for training. The density profiles in our work are mimicking the theoretical density profiles after a long-time evolution on an infinite large lattice. Since our data are generated with numeric simulations, we choose an evolution time which enables the particle's density profile occupying around $80\%$ of the lattice area. This strategy ensures the evolutions  avoid the boundary effects of a finite lattice and in the meanwhile they are good approximations to the long-time evolutions, which means they are time-independent and the minor evolution time changes will not affect our results.

The method to add the noise to our density profiles are different for the data collected in different measurement spaces, i.e. momentum or position. The experimental momentum data  measurement can be implemented in cold atom systems as in Ref.~\cite{Sun_2018, Wu_2016}, where the noise in the data are the shot-noise and Gaussian white noise. The standard deviation of Gaussian noise is set to be $0.02$ in our simulated data, which is a reasonable estimation for current technology based on the error bar ranges in Ref.~\cite{Sun_2018}. The experimental position data measurement can be implemented in cold atom system as in Ref.~\cite{robens2017high} and photonics systems as in Ref.~\cite{xiao2017observation, chen2018observation}  by encoding the position of a walker in either time-bins or spatial modes. The noise in position data includes shot-noise and device noise resulting in the uncertainty in both relative phase and amplitude of the state, which is realised  by the convolution between the perfect state and the point-spread function (PSF) of the system. In our noisy data, the PSF we used is a Gaussian with $0$ as its mean and $2$ as the standard deviation which is also within current experimental techniques level~\cite{Stallinga_2010, Minar}.

 \begin{table*}[h]
\caption{DNN architecture configuration with $LR$ as learning rate.}
\centering
\begin{tabular}{c|c|c|c|c|c|c}
\hline\hline
\multicolumn{7}{c}{Computation Network with $LR=0.0001$} \\
\hline\hline
Block  & Layer & Filter & Size & Activation & Padding & Repetition  \\
\hline
  \multirow{3}{*}{1st} & AvgPool   &    --    & (2,2)  & --   & Valid &  \multirow{3}{*}{2}  \\
                                 & Conv2D   &  8      &(5,5)  &  --   & Valid &  \\
                    & SeparableConv2D & 8  &(5,5)    &ELU  & Same&  \\
\hline
  \multirow{3}{*}{2nd} & AvgPool   &    --    & (2,2)  & --   & Valid &  \multirow{3}{*}{2}  \\
                                 & Conv2D   &  16      &(5,5)  &  --   & Valid &  \\
                    & SeparableConv2D & 16  &(5,5)    &ELU  & Same &  \\
\hline
  \multirow{3}{*}{3rd} & AvgPool   &    --    & (2,2)  & --   & Valid &  \multirow{3}{*}{2}  \\
                                 & Conv2D   &  32      &(5,5)  &  --   & Valid &  \\
                    & SeparableConv2D & 32  &(5,5)    &ELU  & Same &  \\
\hline
4th & Linear & -- & 256 & Relu & --  &1  \\
\hline
5th & Linear &-- & 5  & Softmax & -- & 1 \\
\hline\hline
\multicolumn{7}{c}{Memory Network with $LR=0.4$} \\
\hline\hline
\multicolumn{1}{c|}{Height} & \multicolumn{1}{c|}{Width}&\multicolumn{2}{c|}{Element Size} &\multicolumn{2}{c|}{Decay Factor of $LR$}  &  \multicolumn{1}{c}{Initial Radius}\\
\hline
\multicolumn{1}{c|}{256} & \multicolumn{1}{c|}{256}& \multicolumn{2}{c|}{32} &\multicolumn{2}{c|}{0.9} & \multicolumn{1}{c}{128}\\
\hline

\end{tabular}
\label{table:networkconfig}
\end{table*}

\textit{Identification of perturbed system}.  We consider a perturbation to the Hamiltonian of Eq.~(\ref{Eq:hamiltonian}) given by the addition of a \change{third nearest neighbour (hopping)} term in $x$ direction, which has the expression $\hat{h}^x_\text{N3}=\eta\cos\left(2k_x\right)\sigma_z$, with $k_x\in[-\pi,\pi)$ as the momenta in $x$ direction, and $\eta$ as the perturbation strength. The block-diagonalized   Hamiltonian having \change{a third nearest neighbor (hopping) term}  is $\vec{h}'\cdot\vec{\sigma}=\vec{h}\cdot\vec{\sigma}+\hat{h}^x_\text{N3}$, where $\vec{h}\cdot\vec{\sigma}$ is as in Eq.~(\ref{Eq:hamiltonian}).

All the data in this section is generated in momentum space.   We consider a parameter space for $\vec{h}\cdot\vec{\sigma}$ by fixing $t_{1x}=t_{1y}=1$, $t_2=5$ as before, and with $m\in\left[-20,-10\right)$, $t_3\in\left[-20,20\right]$ corresponding to the area  $\mathcal{C}=\{0, 1\}$  of the phase diagram of the unperturbed system as shown in Fig.~(\ref{phase_diagram}).  For the different  perturbation strength  $\eta=\{3, 6, 9\}$ we generate $196$ data, where $7$ values equally sampled from $m\in\left[-20,-10\right)$ and  $28$ values equally sampled from $t_3\in\left[-20,20\right]$,  where the discretisation resolution for sampling the Hamiltonian parameters is $1.4815$. With the three sets of perturbation data corresponding to $\eta=\{3, 6, 9\}$ as the input, our trained DNN is able to identify their topological phases and phase transitions.  
 The topology identification of the perturbed Hamiltonian is with an averaged accuracy $93.88\%$, with $\{97.96\%, 93.88\%, 89.80\%\}$ for $\eta=\{3, 6, 9\}$ respectively. 

 To show that our trained DNN is detecting the changes to the model caused by the perturbation, we use our trained DNN to track the movement of the topological phase transition as the perturbation is increased.
With the outputs of our trained DNN, we can isolate the location of a topological phase transition as laying between points with different Chern numbers, $(t_{3\text{L}}, t_{3\text{R}})$.  We take the middle point of the two locations $\left(t_{3\text{L}}+t_{3\text{R}}\right)/2$ as the estimate of the phase boundary, and calculate the corresponding shift $\Delta t_3$ from the location of the phase boundary $t_{3|\eta=0}$ in the unperturbed $\eta=0$ model, 
\begin{equation}
\Delta t_3=\left(t_{3\text{L}}+t_{3\text{R}}\right)/2-t_{3|\eta=0}\,.
\end{equation}
We define a \emph{phase boundary shift} $\Delta_\text{DNN}$ for perturbation $\eta$ to be the average of a collection of shifts $\text{sgn}(t_3)\Delta t_3$ with different $m$ along the boundary.  
We note that this method to identify the phase boundary shift is very sensitive to the discretisation resolution of the parameter space on which the DNN is used.  In our simulations, the phase shift affected by the parameters discretisation resolution is $0.7407$, the half of the discretisation resolution. 
 
 Our trained DNN reveals that, while increasing \change{the third nearest neighbor}  coupling strength $\eta$, the phase boundary between $\mathcal{C}=1$ and $\mathcal{C}=0$  shifts outwards (in the direction of increasing magnitude $|t_3|$), i.e., the area of $\mathcal{C}=0$ is increasing with the size of the perturbation.  Specifically, our trained DNN predictions for the phase boundary shifts are $\Delta_\text{DNN}=\left\{1.005, 1.746, 2.686\right\}$ \change{as shown in Fig.~(\ref{boundary_shift})} for $\eta=\{3, 6, 9\}$, which are close to  the theoretical analysis for the corresponding shifts $\Delta=\left\{1, 2, 3\right\}$ indicated from the gapless band structures~\cite{Sticlet_2012}.

\textit{The configurations of our deep neural network for topological phase identification of quantum systems.}  
We use a deep neural network coupled with an external memory for identification of topological phases from the distributions from CTQWs. We take advantage of the most of up-to-date techniques for our computation network design. For the memory network, the simplification of memory operations is achieved by using a  self-organising map (SOM), which is endowed with effective memory addressing and allocation mechanisms. A hybrid learning approach is devised to optimise the network for obtaining promising results.

The detailed architecture and the configuration of our network is  illustrated in Fig.~(\ref{DNN}) and Table~(\ref{table:networkconfig}). There are 6 computation blocks (2 with size $8\times8$, 2 with size $16\times16$ and 2 with size $32\times 32$), 2 fully connected layers and an external memory.   
During the training process, the learning rates ($LR$) for computation network and memory network are $0.0001$ and $0.4$ respectively. The batch size is set as $64$ and the network is trained  $1000$ iterations. 
The learning rate decay factor in our computation network is $0.9$ for  every $100$ iterations. The time constant for SOM is the number of iterations divided by the natural logarithm of initial radius ($128$ in our experiment). 
The labels for memory clusters are probed by tracking the corresponding coordinates  of a few typical data from different topological phases.  The details on the network architecture selection, naive baseline and the misclassified samples interpretation are in  \textit{Appendix}.

Our experiments run on a GPU cluster with three nodes. Each node is with two Intel CPUs of model E5-2680 and 128GB physical memory. For computing acceleration, each CPU manages a separate PCIe slot in which an NVIDIA Quadro P5000 GPU card with 16GB on-board memory installed.

\begin{acknowledgements}
This work is supported by the Australian Research Council via the Centre of Excellence in Engineered Quantum Systems project number CE170100009 and Discovery Project numbers   DP170103073, DP180100670 and DP180100656, and USyd-SJTU Partnership Collaboration Awards. 
The authors acknowledge discussions about noisy experimental data with Wei Sun, Chao Chen, Yu He, and Steven Flammia, Xianmin Jin and comments from Robin Harper and John Manion.
The authors acknowledge the University of Sydney and University of Technology Sydney for providing HPC resources that have contributed to the research results reported in this paper. 
\end{acknowledgements}

{\bf{Data Availability}}

Data are available on request due to privacy or other restrictions.

{\bf{Competing Interests}}

The Authors declare no Competing Financial or Non-Financial Interests.

{\bf{Author Contributions}}

YM and CL designed and performed the experiments. WZ and SB proposed theoretical support. WZ prepared the training data. All authors contributed to writing the paper.


\begin{thebibliography}{99}


\bibitem{Moore_2010}
J.~E. Moore,
\newblock Nature {\bf 464}, 194 (2010).

\bibitem{Hasan_2010}
M.~Z. Hasan and C.~L. Kane,
\newblock Reviews of Modern Physics {\bf 82}, 3045 (2010).

\bibitem{Ryu_2010}
S.~Ryu, A.~P. Schnyder, A.~Furusaki, and A.~W. Ludwig,
\newblock New Journal of Physics {\bf 12}, 065010 (2010).

\bibitem{Qi_2011}
X.-L. Qi and S.-C. Zhang,
\newblock Reviews of Modern Physics {\bf 83}, 1057 (2011).



\bibitem{Li_2014}
C. H. Li et~al.,
\newblock Nature Nanotechnology {\bf 9}, 218 (2014).

\bibitem{Ando_2014}
Y. Ando et~al.,
\newblock Nano Letters {\bf 14}, 6226 (2014).

\bibitem{DC_2018}
M.~DC et~al.,
\newblock Nature Materials {\bf 17}, 800  (2018).

\bibitem{Nayak_2008}
C. Nayak et~al.,
\newblock Reviews of Modern Physics {\bf 80}, 1083 (2008).

\bibitem{Field_2018}
B. Field and T. Simula,
\newblock Quantum Science and Technology {\bf 3}, 045004 (2018).

\bibitem{Wu_2016}
Z.~Wu et~al.,
\newblock Science {\bf 354}, 83 (2016).

\bibitem{Price_2012}
H.~M. Price and N.~R. Cooper,
\newblock  Physical Review A {\bf 85}, 033620 (2012).

\bibitem{Duca_2015}
L.~Duca et~al.,
\newblock Science {\bf 347}, 288 (2015).

\bibitem{Kitagawa_2010}
T.~Kitagawa, M.~S. Rudner, E.~Berg, and E.~Demler,
\newblock Physical Review A {\bf 82}, 033429 (2010).

\bibitem{Kitagawa_2012}
T.~Kitagawa et~al.,
\newblock Nature Communications {\bf 3}, 882 (2012).

\bibitem{Cardano_2016}
F.~Cardano et~al.,
\newblock Nature Communications {\bf 7}, 11439 (2016).

\bibitem{Zhang_2017deco}
W.-W. Zhang, S.~K. Goyal, C.~Simon, and B.~C. Sanders,
\newblock Physical Review A {\bf 95}, 052351 (2017).

\bibitem{Zhang_2017}
W.-W. Zhang, B.~C. Sanders, S.~Apers, S.~K. Goyal, and D.~L. Feder,
\newblock  Physical Review Letters {\bf 119}, 197401 (2017).

\bibitem{Sun_2018}
W.~Sun et~al.,
\newblock  Physical Review Letters {\bf 121}, 250403 (2018).



\bibitem{zhang2017detecting}
X. Zhan et~al.,
\newblock  Physical Review Letters {\bf 119}, 130501 (2017).

 
\bibitem{Sticlet_2013}
D.~Sticlet, and F.~Pi{\'e}chon,
\newblock Physical Review B {\bf 87}, 115402 (2013).
\bibitem{Montambaux2012}
G. Montambaux,
\newblock The European Physical Journal B {\bf 85}, 375 (2012).
 




\bibitem{Venegas_2012}
S.~E. Venegas-Andraca,
\newblock Quantum Information Processing {\bf 11}, 1015 (2012).

\bibitem{portugal2013quantum}
R.~Portugal,
\newblock Quantum walks and search algorithms,
\newblock New York: Springer, (2013).

\bibitem{Flurin_PRX_2017}
E.~Flurin et~al.,
\newblock  Physical Review X {\bf 7}, 031023 (2017).

\bibitem{xiao2017observation}
L.~Xiao et~al.,
\newblock Nature Physics {\bf 13}, 1117 (2017).

\bibitem{Sticlet_2012}
D.~Sticlet, F.~Pi{\'e}chon, J.-N. Fuchs, P.~Kalugin, and P.~Simon,
\newblock Physical Review B {\bf 85}, 165456 (2012).

\bibitem{Asboth_2016}
J.~K. Asb{\'o}th, L.~Oroszl{\'a}ny, and A.~P{\'a}lyi,
\newblock Lecture Notes in Physics {\bf 919} (2016).


\bibitem{Schmidt_2009}
M.~Schmidt and H.~Lipson,
\newblock Science {\bf 324}, 81 (2009).



\bibitem{goodfellow2016deep}
I.~Goodfellow, Y.~Bengio and A.~Courville, 
\newblock Deep learning,
\newblock MIT press, (2016).


\bibitem{krizhevsky2012imagenet}
A.~Krizhevsky, I.~Sutskever, and G.~E. Hinton,
\newblock Imagenet classification with deep convolutional neural networks,
\newblock  Advances in Neural Information Processing Systems, 
  1097, (2012).

\bibitem{esteva2017dermatologist}
A.~Esteva et~al.,
\newblock Nature {\bf 542}, 115 (2017).

\bibitem{shallue2018identifying}
C.~J. Shallue and A.~Vanderburg,
\newblock The Astronomical Journal {\bf 155}, 94 (2018).

\bibitem{Cai_2015}
X.-D. Cai et~al.,
\newblock  Physical Review Letters {\bf 114}, 110504 (2015).

\bibitem{Schuld_2015}
M.~Schuld, I.~Sinayskiy, and F.~Petruccione,
\newblock Contemporary Physics {\bf 56}, 172 (2015).

\bibitem{Dunjko_2016}
V.~Dunjko, J.~M. Taylor, and H.~J. Briegel,
\newblock  Physical Review Letters {\bf 117}, 130501 (2016).

\bibitem{Biamonte_2017}
J.~Biamonte et~al.,
\newblock Nature {\bf 549}, 195 (2017).

\bibitem{Mott_2017}
A.~Mott, J.~Job, J.-R. Vlimant, D.~Lidar, and M.~Spiropulu,
\newblock Nature {\bf 550}, 375 (2017).

\bibitem{Broecker_2017}
P.~Broecker, J.~Carrasquilla, R.~G. Melko, and S.~Trebst,
\newblock Scientific Reports {\bf 7}, 8823 (2017).

\bibitem{Carrasquilla_2017}
J.~Carrasquilla and R.~G. Melko,
\newblock Nature Physics {\bf 13}, 431 (2017).


\bibitem{Zhang_2017qml}
Y.~Zhang, and E.-A.~Kim,
\newblock  Physical Review Letters {\bf 118}, 216401 (2017).


\bibitem{Zhang_2018}
P.~Zhang, H.~Shen, and H.~Zhai,
\newblock  Physical Review Letters {\bf 120}, 066401 (2018).

\bibitem{Choo_2018}
K. Choo, G. Carleo, N. Regnault and T. Neupert,
\newblock  Physical Review Letters {\bf 121}, 167204 (2018).

\bibitem{Lu_2018}
S. Lu et~al.,
\newblock  Physical Review A {\bf 98}, 012315 (2018).

\bibitem{arrazola2018machine}
Arrazola, Juan Miguel et~al.,
\newblock Quantum Science and Technology {\bf 4}, 024004 (2019).


\bibitem{Caio_2019}
M.~Caio, M.~Caccin, P.~Baireuther, T.~Hyart and M.~Fruchart,
\newblock arXiv:1901.03346 (2019).
 
\bibitem{Mehta2019}
P.~Mehta et~al.,
\newblock arXiv:1803.08823 (2019).
\bibitem{Rem_2018}
B.S.~Rem et~al.,
\newblock  arXiv:1809.05519 (2018).
\bibitem{Sarma_2019}
S. D.~Sarma, S.-L. Deng and L.-M., Duan,
\newblock Physics Today  {\bf{72}}, 48 (2019).
\bibitem{Schuld_2019}
M.~Schuld,
\newblock Nature {\bf{567}}, 179 (2019).
\bibitem{Rodriguez2019}
J.~F.~Rodriguez-Nieva and M.~S.~Scheurer, 
\newblock Doi:10.1038/s41567-019-0512-x,
\newblock Nature Physics (2019).
 
\bibitem{hubel2012david}
D.~Hubel and T.~Wiesel,
\newblock Neuron {\bf 75}, 182 (2012).

\bibitem{lecun2015deep}
Y.~LeCun, Y.~Bengio, and G.~Hinton,
\newblock Nature {\bf 521}, 436 (2015).



\bibitem{Graves_2016}
A.~Graves et~al.,
\newblock Nature {\bf 538}, 471 (2016).

\bibitem{mclaughlin2015data}
N.~McLaughlin, J.~M. Del~Rincon, and P.~Miller,
\newblock Data-augmentation for reducing dataset bias in person re-identification,
\newblock  2015 12th IEEE International Conference on Advanced Video and Signal Based Surveillance (AVSS), IEEE 1, (2015).

\bibitem{crispell2017dataset}
D.~Crispell, O.~Biris, N.~Crosswhite, J.~Byrne, and J.~L. Mundy,
\newblock  arXiv:1704.04326  (2017).

\bibitem{devries2017dataset}
T.~DeVries and G.~W. Taylor,
\newblock arXiv:1702.05538  (2017).

\bibitem{chen2018observation}
C.~Chen et~al.,
\newblock Physical Review Letters {\bf 121}, 100502 (2018).

\bibitem{robens2017high}
C.~Robens et~al.,
\newblock Optics Letters {\bf 42}, 1043 (2017).

\bibitem{xian2018zero}
Y. Xian, C. Lampert, B. Schiele, and Z. Akata,
\newblock Zero-shot learning-a comprehensive evaluation of the good, the bad and the ugly, 
\newblock  IEEE Transactions on Pattern Analysis and Machine Intelligence,  (2018).

\bibitem{Stallinga_2010}
S. Stallinga and B. Rieger,
\newblock Optical Express {\bf 18}, 24461 (2010).

\bibitem{Minar}
J. Min\'a\ifmmode \check{r}\else \v{r}\fi{}  et~al.,
\newblock Physical Review A {\bf 77}, 052325 (2008).

\end{thebibliography}

\clearpage
\newpage

\appendix
\renewcommand{\thefigure}{S\arabic{figure}}
\renewcommand{\thetable}{S\arabic{table}}
\setcounter{figure}{0}
\setcounter{table}{0}
 
\section*{Appendix}
{\bf{\textit{Neural architecture evaluation and naive baseline.}}} 
\change{The incorporating of a self-organising map (SOM) module into DNN as proposed in this paper, requires special considerations from implementing perspective to mitigate the complexity.  From the pragmatic perspective, a simpler feature-maps from the final convolutional layer are preferred. To evaluate the neural architecture, we empirically choose two other types of network structures, aka multi-layer perceptron (MLP) and vanilla CNN, and benchmark results produced by DNN  to the MLP and vanilla CNN  to guarantee the high accuracy of prediction for quantum topology identification. The configurations  of MLP and vanilla CNN we constructed for comparison are shown in Table~\ref{MLPCNN}.}

The data in this work is generated via simulation where  it is straightforward to set the size of data for different categories in a similar scale to ensure a balanced case. We note that data from the real experiment could potentially be in an imbalanced situation. Considering the imbalanced data may lead to the mis-behaviour of networks, an overall prediction accuracy based on cross-entropy cannot be used for optimisation in this circumstance.  
Hence, other indicators such as precision and recall are used for optimisation with imbalanced data. To investigate  of the  performance of our method with the imbalanced data, maintaining the individual accuracy for each class is quite insightful  for a rough estimation of performance drop.

The corresponding performance comparisons with the individual and averaged accuracies for different neural architectures
including a multilayer perceptron (MLP), Vanilla CNN, and our DNN, are shown in Table~\ref{perform_diff_net}. 
For position data, we have observed it is difficult for any neural network structure to effectively distinguish between the $\mathcal{C}=\pm2$ categories, hence only the average accuracy of  the sets $\mathcal{C}=0,\pm1$ are considered for the network benchmark. For momentum data, there are accuracy improvement with the complexity of the network.  In addition, DNN is known to be robust with the ill-distributed data. All of these considerations lead to our final choice of the current network structure.

\begin{figure*}[h] 
\includegraphics[width=0.88\textwidth]{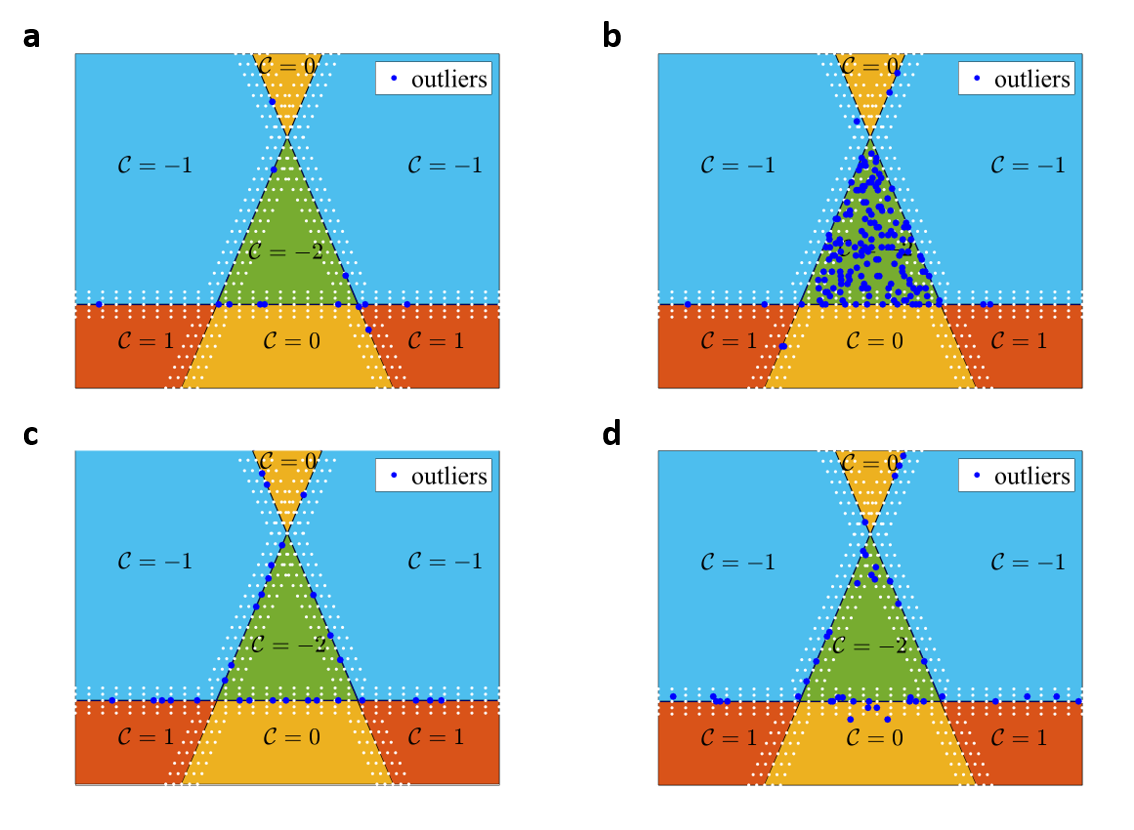}
\caption{The location of the misclassified samples (outliers) for whole momentum data (a) and whole position data (b), transition momentum data (c) and transition position data (d). While all the misclassified samples for the test data from whole momentum, transition momentum and transition position sets are located along the phase boundaries in the phase diagram as expected, the misclassified samples for the test data from position  whole  region locate along the phase boundaries in the phase diagram and the $\mathcal{C}=|2|$ phase area.}
\label{misclassification}
\end{figure*}

{\bf{\textit{Misclassified sample interpretation.}}}
Although the interpretation for the dynamics and representations of the neural network is a challenge in general, we can nonetheless attempt to illustrate the results of misclassification with the aim of giving some hints about the behaviour of the network.  We demonstrate the misclassified samples of whole momentum and  position data, and the transition momentum and position data in Fig.~\ref{misclassification}(a)-(d) respectively.
We see that the misclassified samples tend to correspond to points near a phase transition for whole momentum, the transition momentum and position data.  However, for whole position data the DNN is not able to identify $\mathcal{C}=\pm2$ which results the big misclassification cluster located at the $\mathcal{C}=\pm2$ area as shown in Fig.~\ref{misclassification}(b), which is also indicated by the corresponding accuracies of $70.3\%$ and $24.1\%$ corresponding to $\mathcal{C}=-2$ and $\mathcal{C}=2$.

\begin{table*}[h]
\change{
\caption{Architecture configurations of the multi-layer perceptron (MLP) and vanilla CNN.}
\centering
\begin{tabular}{c|c|c|c|c|c}
\hline\hline
\multicolumn{6}{c}{Multi-layer perceptron (MLP)} \\
\hline\hline
Block  & Layer & Filter & Size & Activation & Padding   \\
\hline
--   & AvgPool  & -- & (2, 2) & -- & --    \\
\hline
--   & Linear & -- & 2048 & ELU & --    \\
\hline
--   & Linear & -- & 256 & ELU & --    \\
\hline
--   & Linear &-- & 5   & Softmax & -- \\
\hline\hline
\multicolumn{6}{c}{Vanilla CNN} \\
\hline\hline
Block  & Layer & Filter & Size & Activation & Padding   \\
\hline
1st  & AvgPool  & -- & (2, 2) & -- & Same   \\
\hline
  \multirow{2}{*}{2nd} 
                               & Conv2D   &  32      &(5,5) &  ELU  &   \multirow{2}{*}{Same}  \\
                                  & AvgPool   &    --    & (2,2)  &  --   &  \\
\hline
  \multirow{2}{*}{3rd} 
                               & Conv2D   &  64      &(5,5) &  ELU  &   \multirow{2}{*}{Same}  \\
                                  & AvgPool   &    --    & (2,2)  &  --   &  \\
\hline
4th & Linear & -- & 256 & ELU & --    \\
\hline
5th & Linear &-- & 5  & Softmax & --  \\
\hline
\end{tabular}
\label{MLPCNN}}
\end{table*}

\begin{table*}[h]
\caption{\change{Performance of different Neural Network models} including MLP network, Vanilla CNN with 2 convolutional layers, DNN with 6 convolutional blocks, and the number of parameters in the corresponding Neural Networks. Here all the data are obtained with the ideal data without any noise. (Note: if the data denoted with an asterisk (*) for the position data sets $\mathcal{C}=\pm2$ are excluded, we obtain a higher overall accuracy as denoted by a dagger ($^\dagger$) compared with the overall accuracy including this data.)}
\centering
\begin{tabular}{c|c|c|c|c|c|c|c}
\hline\hline
 \multicolumn{8}{c}{The statistical accuracy obtained with MLP network}\\
\hline\hline
 \multicolumn{2}{c}{Density Profile Data} &  \multicolumn{5}{|c|}{$\mathcal{C}$}  & \multirow{2}{*}{Overall} \\
\cline{1-7}
Phase Diagram Area  & Measurement  Domain & $-2$ & $-1$ & $0$ & $1$ & $2$  &  \\
\hline
  \multirow{2}{*}{Whole} & Momentum & 0.904   &  0.961   & 0.960  &0.941   &  1.000 & 0.953  \\
                                    & Position        &  0.867* & 0.907     &  0.947  & 0.968  & 0.095*  & 0.760 ($0.941 ^\dagger$) \\
\hline\hline
 \multicolumn{8}{c}{The statistical accuracy obtained with Vanilla CNN (2 conv layers)}\\
\hline\hline
 \multicolumn{2}{c}{Density Profile Data} &  \multicolumn{5}{|c|}{$\mathcal{C}$}  & \multirow{2}{*}{Overall}\\
\cline{1-7}
Phase Diagram Area  & Measurement  Domain & $-2$ & $-1$ & $0$ & $1$ & $2$  & \\
\hline
  \multirow{2}{*}{Whole} & Momentum &0.944  &0.966  &0.971  & 0.966   & 1.000   & 0.969\\
                                    & Position & 0.457*  &0.944  &0.946   &0.984   &0.339*   & 0.738  ($0.958^\dagger$) \\
\hline\hline
 \multicolumn{8}{c}{The statistical accuracy obtained with DNN (6 conv layers)}\\
\hline\hline
 \multicolumn{2}{c}{Density Profile Data} &  \multicolumn{5}{|c|}{$\mathcal{C}$}  & \multirow{2}{*}{Overall}\\
\cline{1-7}
Phase Diagram Area  & Measurement  Domain & $-2$ & $-1$ & $0$ & $1$ & $2$  &    \\
\hline
  \multirow{2}{*}{Whole} & Momentum &0.974  & 0.961 &0.969  & 0.969  &1.000  & 0.974  \\
                                    & Position & 0.703* & 0.935 & 0.937  &0.975   &0.241*  & 0.761 ($0.949^\dagger$) \\
\hline
\end{tabular}
\label{perform_diff_net}
\end{table*}

\end{document}